\documentclass[%
a4paper,
superscriptaddress,
notitlepage,
 amsmath,amssymb,
 aps,
 twocolumn,
 prl
 floatfix
]{revtex4-1}
\usepackage{centernot}
\usepackage{tikz}
\usetikzlibrary{calc,arrows,decorations.pathreplacing,backgrounds}
\colorlet{mylinkcolor}{blue!66!black!80}
\usepackage{mathtools,amssymb,amsfonts}
\usepackage[colorlinks=true,linkcolor=mylinkcolor,citecolor=mylinkcolor,filecolor=cyan,urlcolor=mylinkcolor,breaklinks=true]{hyperref}
\usepackage{xcolor}

\usepackage[utf8]{inputenc}
\usepackage{booktabs}
\usepackage{bbm,mathrsfs,mathtools}
\usepackage{graphicx}

\DeclareSymbolFontAlphabet{\mathscrsfs}{rsfs}

\begin{document}
\title{Milestoning estimators of dissipation in systems
  observed at a coarse resolution: When ignorance is truly bliss}

\author{Kristian Blom }
% \email{kblom@mpinat.mpg.de}
\affiliation{Mathematical bioPhysics group, Max Planck Institute for Multidisciplinary Sciences, G\"{o}ttingen 37077, Germany}

\author{Kevin Song}
\affiliation{Department of Computer Science, University of Texas at Austin, Austin, Texas 78712, USA}

\author{Etienne Vouga}
\affiliation{Department of Computer Science, University of Texas at Austin, Austin, Texas 78712, USA}

\author{Alja\v{z} Godec}
\email{agodec@mpinat.mpg.de}
\affiliation{Mathematical bioPhysics group, Max Planck Institute for Multidisciplinary Sciences, G\"{o}ttingen 37077, Germany}

\author{Dmitrii E. Makarov}
\email{makarov@cm.utexas.edu}
\affiliation{Department of Chemistry and Oden
Institute for Computational Engineering and Sciences, The University
of Texas at Austin, Austin, Texas 78712, USA}
%\author{\red{Horse Wolfgang}}
%\affiliation{Department of Chemistry and Oden
%Institute for Computational Engineering and Sciences, The University of Texas at Austin, Austin, Texas 78712, USA}
\date{\today}

\date{\today}

\begin{abstract}
Many non-equilibrium, active processes are observed at a coarse-grained level,
where different microscopic configurations are projected onto the
same observable state. Such ``lumped'' observables
display memory, and in many cases the irreversible character of the
underlying microscopic dynamics becomes blurred, e.g.,\ when the
projection hides dissipative cycles. As a result, the
observations appear less irreversible, and it is very challenging to
infer the degree of broken time-reversal symmetry. Here we show,
contrary to intuition, that by ignoring parts of the already coarse-grained
  state space we
  may ---via a process called milestoning--- %quite generally 
  improve
  entropy-production estimates. Milestoning systematically renders  observations
"closer to underlying microscopic dynamics" and
%  ``more Markovian'' and
  thereby improves thermodynamic inference from lumped data assuming a given range of memory. Moreover, whereas the correct general physical definition of time-reversal in the presence of memory remains unknown, we here show by means of systematic,  physically relevant examples that at least for semi-Markov processes of first and second order, waiting-time contributions arising from adopting a naive Markovian definition of time-reversal generally must be discarded.   
\end{abstract}
\maketitle

\noindent \textbf{One-sentence summary:~}Non-equilibrium processes are often observed at a coarse resolution, blurring the irreversibility of dynamics; contrary to intuition, we show that systematically increasing ignorance via a process called milestoning %quite generally 
can improve thermodynamic inference. 

%\textit{Introduction.---}
\section*{Introduction}
Although the nonequilibrium character of
%living organisms
active systems
may often be self-evident (e.g.,\ the directed
motion of a kinesin motor or persistent rotation of the
$\rm F_{1}$-ATPase \cite{yasuda2001resolution}), differentiating in general between equilibrium and
non-equilibrium steady states at mesoscopic and microscopic scales is
surprisingly difficult \cite{doi:10.1021/acs.jpclett.2c03244,
  ratzke2014four, tu2008nonequilibrium, gnesotto2018broken,
  li2019quantifying}. A key reason is that a broken
  time-reversal symmetry often emerges on length scales where
  detailed microscopic
  information is unavailable, e.g.,\ below the diffraction
  limit. In particular, single-molecule experiments
probe inherently coarse-grained representations of microscopic dynamics
\cite{makarov2015single} and may thus underestimate
\cite{doi:10.1021/acs.jpclett.2c03244} or even ``be blind to'' relevant slow dissipative
degrees of freedom \cite{PhysRevLett.108.220601}.

Coarse-graining may be a consequence of experimental limitations as
well as of data analysis: for example, the location of a microscopic
probe is, most optimistically, limited by the pixel size, which
introduces spatial coarse-graining \cite{Need}. Similarly, when protein folding is probed using optical tweezers
\cite{hoffer2019probing} or fluorescence resonance energy transfer \cite{doi:10.1021/acs.jpclett.2c03244, SCHULER200816, CHUNG201830}, many molecular conformations with, respectively, the
same extension or with the same 
distance between fluorescent probes are lumped
onto a single state. This projects the high-dimensional conformational
dynamics %of the molecule
onto a single lumped coordinate. 
\begin{figure}[t!]
    \centering
    \includegraphics[width=0.48\textwidth]{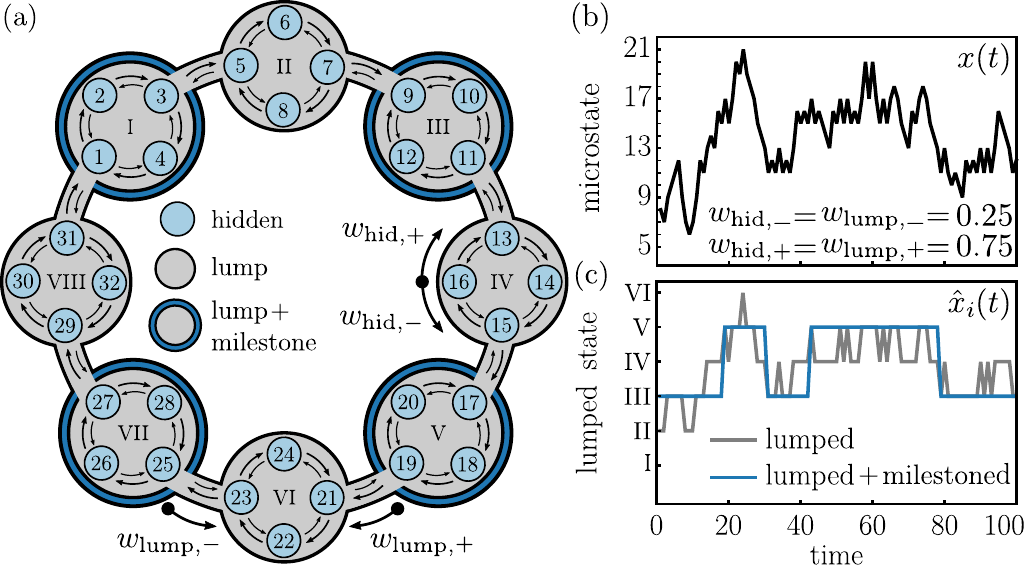}
    \caption{\textbf{Lumping and milestoning a discrete-state Markov
        chain.} (a) Schematic of a discrete-state Markov chain with 32 hidden
      microstates lumped into domains I-VIII
      (gray) each containing 4 microstates. Such a partitioning arises
      naturally when the underlying process is observed with coarse resolution. Milestoning involves further post-processing of the lumped process: some
      lumped states (here, the domains I, III, V, and VII)
      are declared to be milestones (blue). Once the trajectory enters a
      milestone, the coarse-grained state of the  milestone
      trajectory remains in said milestone until it enters a different milestone. (b) Trajectory
      of the full Markov chain $x(t)$. (c) Trajectory of the
      coarse-grained semi-Markov chain $\hat{x}_{i}(t)$ obtained by lumping
      (gray) and further milestoning (blue).
     }
    \label{lumping}
\end{figure}

Consider the total steady-state entropy-production rate
(i.e.,\ dissipation rate), estimated as \cite{peliti2021stochastic, seifert2012stochastic}
\begin{eqnarray}
   \langle \dot{S}[x] \rangle \equiv
  \lim\limits_{t\rightarrow\infty}\frac{1}{t}
  \left\langle\ln{\left(\frac{P[\{x(\tau)\}_{0\le \tau\le
          t}]}{P[\{\theta x(\tau)\}_{0\le \tau\le t}]}\right)} \right\rangle,
    \label{S}
\end{eqnarray}
where $x(\tau)$ denotes the system’s %microscopic 
state at time
  $\tau$, $P[\{x(\tau)\}_{0\le \tau\le t}]$ is the probability of a
forward path $\{x(\tau)\}_{0\le \tau\le t}$ and $\theta$ denotes the
\emph{physically consistent} time-reversal operation, such that
$P[\{\theta x(\tau)\}_{0\le
    \tau\le t}]$ corresponds to the probability of the
%\emph{physical}
time-reversed path. In Eq.~\eqref{S} Boltzmann’s
constant was set to 1, $\langle\cdot\rangle$ indicates averaging
over $P[\{x(\tau)\}_{0\le \tau\le t}]$, and we tacitly
assumed that the dynamics is ergodic. When $x(\tau)$ is a Markov
process, $\{\theta x(\tau)\}_{0\le \tau\le t}=\{\epsilon
x(t-\tau)\}_{0\le \tau\le t}$ where $\epsilon$ denotes that we must
simultaneously change the sign of all degrees of freedom that are
\emph{odd} under time reversal (such as
momenta).   

Note that in the presence of memory, the physically consistent form of time-reversal
$\theta$ is generally \emph{not} known. In particular, even
  in the absence of momenta  
  one \emph{cannot} simply assume the Markov definition, i.e.,\ $\{\theta x(\tau)\}_{0\le
    \tau\le t}\ne\{x(t-\tau)\}_{0\le \tau\le t}$ (see
  \cite{Qian,PhysRevX.11.041047,comment} and counterexamples below).
Memory effects
require adapted definitions of time reversal, analogous to momenta in
inertial systems \cite{PhysRevX.11.041047,comment}. In particular, naively
setting $\{\theta x(\tau)\}_{0\le
    \tau\le t}{\overset{!}{=}}\{x(t-\tau)\}_{0\le \tau\le t}$ in Eq.~\eqref{S} may yield a positive entropy production for
coarse-grained representations of manifestly reversible microscopic
dynamics, as it leads to a "waiting-time contribution" to dissipation \cite{Qian,PhysRevX.11.041047,comment}.  

If the time-reversal operation and coarse-graining are carried
  out correctly, projected representations of microscopic dynamics
are usually ``less irreversible'' than the true dynamics
\cite{Massi,PhysRevX.11.041047,van2022thermodynamic,Polettini,Snippets, deguenther2023fluctuating}.  That is, if %instead %of $x(t)$
a coarse-grained
representation $\hat{x}(t)$ of the full dynamics is considered, we typically
expect to underestimate  \footnote{In case of a perfect time-scale
separation between the observed and hidden degrees of freedom, the total entropy production is not necessarily
underestimated by a coarse-grained representation.} the true
entropy production \cite{Massi,PhysRevX.11.041047,yu2021inverse, van2022thermodynamic}, i.e.
\begin{equation}
    0\leq \langle \dot{S}[\hat{x}] \rangle \leq \langle \dot{S}[x]
    \rangle.
\label{ineq}    
\end{equation}
\begin{figure*}[ht!!]
    \centering
    \includegraphics[width=0.99\textwidth]{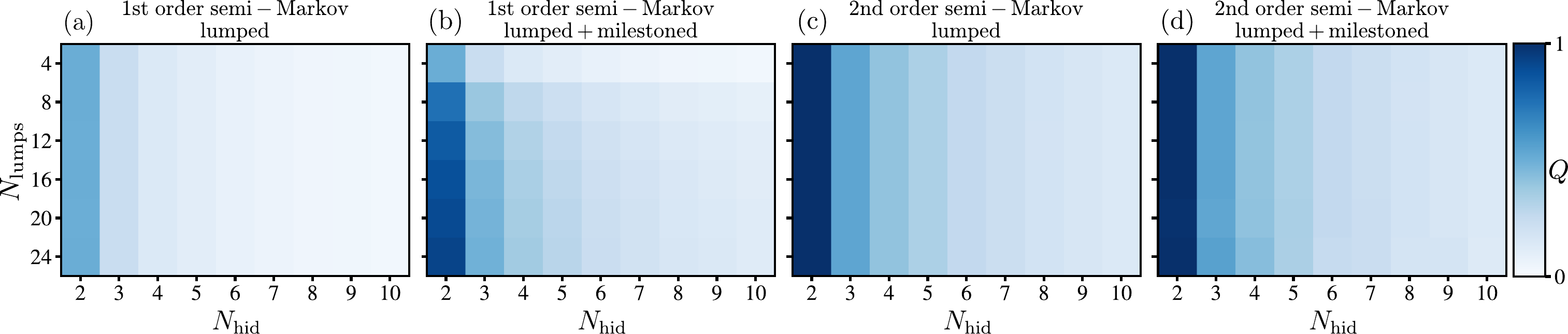}
    \caption{\textbf{Estimating the entropy production from a
        coarse-grained trajectory with hidden cycles.} The diffusive
      trajectory is governed by the discrete-state ring-like Markov
      chain shown in Fig.~\ref{lumping}, with $N_{\rm lumps}$ the
      number of lumped states, $N_{\rm hid}$ the number of hidden
      microstates per lumped state, transition rates
      $\{w_{{\rm hid},+},w_{{\rm hid},-}\}=\{w_{{\rm lump},+},w_{{\rm lump},-}\}=\{0.6,0.4\}$, and $10^{8}$ discrete steps. (a, c) Quality factor $Q$ given
      by Eq.~\eqref{Q} for the entropy production estimated with the
      lumped trajectory assuming $1^{\rm st}$ order semi-Markov (see Eq.~\eqref{Ssemi1}) (a) and $2^{\rm nd}$ order semi-Markov (see Eq.~\eqref{Ssemi}) (c). (b, d) Quality factor for the entropy production
      estimated with the post-lumped milestoned trajectory with 4 equidistant milestones, assuming $1^{\rm st}$
      order semi-Markov (b) and $2^{\rm nd}$ order semi-Markov (d).}
    \label{ringresults}
\end{figure*}
Recently, there has been a surge in interest in thermodynamic inference from
coarse-grained, partially observed dynamics \cite{Puglisi,Teza,Talkner, van2022thermodynamic, PhysRevX.11.041047, Polettini, deguenther2023fluctuating}.  
The arguably most direct method to infer a lower bound on the
entropy production %in a system observed at any coarse resolution
is via the 
thermodynamic uncertainty relation (TUR)
\cite{PhysRevLett.114.158101,Horowitz2019NP,Gingrich2017JPAMT,Vu2020PRE,Manikandan2020PRL,Otsubo2020PRE,Li2019NC,Koyuk2021JPAMT,Need,Dieball2022PRR,Dechant2021PRR,Dechant2021PRX,PhysRevLett.130.087101,cost}. Moreover,
approaches were developed that exploit the information encoded in the non-Markovian character
to infer bounds on dissipation \cite{martinez2019inferring, Ehrich_2021, PhysRevLett.127.198101, Snippets, deguenther2023fluctuating}.

Given the inequality in Eq.~\eqref{ineq}, optimal estimation of entropy production from experimental observations remains an open question. Indeed, different coarse-grained representations yield different estimates; moreover, in any realistic
experimental setting neither the microscopic dynamics nor the precise
coarse-graining is known. Recently, it
  was found that a coarse-graining method called ``milestoning'' 
preserves the microscopic entropy production even in the absence of a
time-scale separation between hidden and observed degrees of freedom, if no dissipative cycles
are hidden \cite{PhysRevX.11.041047,hartich2021violation}. In 
contrast, the more common coarse-graining approach called ``lumping''
(see Fig.~\ref{lumping} and \cite{Massi}) in general may underestimate
$\langle \dot{S}[x] \rangle$ even in the presence of such time-scale
separation \cite{hartich2021violation}. Milestoning, originally developed as a
method for efficient computation \cite{bello2015exact,
  elber2016perspective, elber2020molecular}, projects the dynamics
onto ``hopping'' between a set of milestones that do not cover the
entire configuration space, thereby mapping continuous or
discrete-state ``microscopic'' dynamics onto a generally
non-Markovian random walk.
Intriguingly, we here show that milestoning, if used to post-process
lumped dynamics, can improve thermodynamic inference. In other words, milestone trajectories, obtained by \emph{discarding} certain details of lumped trajectories (as in Fig.~\ref{lumping}), can provide improved estimates of dissipation; as such, milestoning analysis is directly applicable to experimentally observed dynamics, which are inherently lumped.  

%\indent

Here we focus on the direct approach~\eqref{S} %Letter (not letter if we want to submit to P'NIS)
and demonstrate that, contrary to naive intuition, milestoning already
lumped dynamics
  may improve entropy-production estimates via
  Eq.~\eqref{S} \emph{given the same coarse-grained
  trajectories and even in the presence of hidden cycles}. That is, by ignoring parts of the coarse-grained
  state space and thus introducing additional ``controlled ignorance''
  we actually render the trajectory ``closer to the underlying microscopic dynamics'' and
  thereby improve the dissipation estimator~\eqref{S} \cite{10.1063/1.3590108}. We explicitly
  address the scenario where lumping hides dissipative cycles and
  compare the milestoning estimate with the bound inferred from the
  TUR. We
  show how post-lumped milestoning may be used systematically to improve
  thermodynamic inference, i.e.,\ to enhance the precision of estimating dissipation from coarse observations. We stress the importance of
considering a  physically consistent time-reversal operation in the presence of memory, and show that
%(spurious) 
asymmetric waiting-time contributions that emerge upon milestoning generally do not contribute to dissipation.      
 
%\emph{Milestoning dynamics with hidden cycles.---}
\section*{Milestoning dynamics with hidden cycles}
We begin with a 
 toy model that contains hidden dissipative cycles (see \cite{PhysRevE.105.044113, PhysRevLett.125.110601, nitzan2022universal} for similar toy models). Models of this kind can emerge as descriptions of a system of magnetically coupled colloidal particles \cite{Uhl_2018, PhysRevLett.108.220601}, chemical reaction networks \cite{10.1063/1.4935064}, circular hidden Markov models \cite{CHOO200484, 10.1093/bioinformatics/18.4.617}, and molecular conformation dynamics \cite{ratzke2014four}. Here we consider 
 a discrete-time Markov process as shown in Fig.~\ref{lumping}a. The observed dynamics involves $N_{\rm lump}$ observable states arranged along a ring, and each such observable state contains $N_{\rm hid}$ microscopic states forming a ring  (see
  Fig.~\ref{lumping}a). Within each lump there are two maximally separated microstates which are connected to the previous/next lump (e.g.,~microstate 3 and 5 in Fig.~\ref{lumping}a), and in the SM we also consider the case with three microstates connected between the lumps. The transition probabilities for internal microstates (i.e.,~microstates not connecting two lumps) are $w_{{\rm hid},+}$ in the clockwise direction, and $w_{{\rm hid},-}$ in the counterclockwise direction. For microstates connecting two lumps, we have $w_{{\rm lump},\pm}$ for transitions between the lumps, and $(1-w_{{\rm lump},\pm})w_{{\rm hid},\pm}$ for internal transitions.  Thus, for $w_{{\rm hid},+} \neq w_{{\rm hid},-}$, lumping the $N_{\rm hid}$ states into a single observable state hides
  dissipative cycles when $N_{\rm hid}>2$. A microscopic and lumped trajectory are shown in
  Fig.~\ref{lumping}b and c (gray line), respectively. We now
  milestone the trajectory by taking 4 equidistant lumps as milestones
  (see Fig.~\ref{lumping}a and Fig.~\ref{lumping}c  blue line).
  
To evaluate $\langle \dot{S}[\hat{x}_{i}]\rangle$, where the subscript $i=\{ \rm lump, mil \}$ indicates whether the entropy production is estimated for the lumped trajectory or the milestone trajectory, we use Eq.~\eqref{S} and measure lump/milestone-sequence frequencies in the trajectory \cite{roldan2010estimating, roldan2012entropy}. In the presence of
long-range memory this is a computationally demanding task requiring many independent steady-state trajectories or one extremely long trajectory. Even if this is given, a finite range of memory must be assumed in practical computations of path probabilities $P[\{x(\tau)\}_{0\le \tau\le t}]$ and $P[\{\theta x(\tau)\}_{0\le \tau\le t}]$. A numerical
implementation of the inference is provided in \cite{Blom2023}. 

Note that $\hat{x}_i(t)$ in
Fig.~\ref{lumping}c is indeed a non-Markovian process. This
is manifested in the fact that the estimate of $\langle \dot{S}[\hat{x}_i] \rangle$ 
strongly depends on the assumption of the underlying extent of 
memory. That is, 
considering only the preceding step---in the $1^{\rm st}$ order semi-Markov approximation---we assume that
$\hat{x}_i(t)$ depends only on $\hat{x}_i(t-1)$ 
but not on
$\hat{x}_i(t-k)$ for $k\geq 2$, leading to the one-step affinity estimate (assuming all coarse states and milestones are equivalent as in Fig.~\ref{lumping}a)
\begin{equation}
\langle \dot{S}^{\rm aff}_{1}[\hat{x}_{i}]\rangle = \frac{p^{i}_{+}-p^{i}_{-}}{\langle \tau_{i}\rangle }\ln{\left(\frac{p^{i}_{+}}{p^{i}_{-}}\right)},
    \label{Ssemi1}
\end{equation}
where $\langle \tau_{i} \rangle$ is the average waiting time within a coarse-grained
state, and $p^{i}_{\pm}$ are the forward/clockwise ($+$) and backward/counterclockwise
($-$) jump probabilities in the sequence of \emph{distinct}
coarse-grained states (i.e.,\ upon removing repeated consecutive coarse-grained states).

In the $k^{\rm th}$ order semi-Markov approximation we assume that the probability of $\hat{x}_i(t)$
depends on the sequence $\{\hat{x}_i(t-1),\ldots,\hat{x}_i(t-k)\}$. To be concrete, $\hat{x}_{i}(t)$ in
  Fig.~\ref{lumping}c is a $2^{\rm nd}$ order semi-Markov
  process \cite{martinez2019inferring} (see also \cite{Ehrich_2021,
    PhysRevLett.127.198101}), where the jump probabilities \emph{and} waiting times depend on the
  previous state. Naively, taking $\{\theta \hat{x}(\tau)\}_{0\le
    \tau\le t}{\overset{!}{=}}\{\hat{x}(t-\tau)\}_{0\le \tau\le t}$ in Eq.~\eqref{S}, yields two
  contributions: the two-step affinity contribution $\langle
  \dot{S}^{\rm aff}_{2}[\hat{x}_{i}]\rangle$, and a contribution of waiting
  times $\langle\dot{S}^{\rm  wt}_2[\hat{x_{i}}]\rangle$ \cite{martinez2019inferring}. In particular, for $\hat{x}_{i}(t)$ in
  Fig.~\ref{lumping}c the two-step affinity contribution reads \cite{martinez2019inferring, hartich2021violation}
\begin{equation}
\langle \dot{S}^{\rm aff}_{2}[\hat{x}_{i}]\rangle = \frac{p^{i}_{+}-p^{i}_{-}}{\langle \tau_{i}\rangle }\ln{\left(\frac{\phi^{i}_{++}}{\phi^{i}_{--}}\right)},
    \label{Ssemi}
\end{equation}
where $\phi^{i}_{\pm\pm}$ are conditional splitting
probabilities of making a forward/backward jump, given that the previous jump occurred in the forward/backward direction. The explicit form of the waiting-time contribution $\langle
  \dot{S}^{\rm wt}_{2}[\hat{x}_{i}]\rangle$ (see Eq.~\eqref{wait_cont} below) suggests that as soon as waiting time distributions to make a forward versus a backward jump (conditioned on that the preceding jump has also occurred in the forward or backward direction, respectively) are not equal, these contribute to dissipation \cite{martinez2019inferring}. Below, however, we give examples where $\langle
  \dot{S}^{\rm wt}_{2}[\hat{x}_{i}]\rangle>0$ for systems obeying detailed balance; therefore, the naive Markovian time-reversal operation leading to $\langle
  \dot{S}^{\rm wt}_{2}[\hat{x}_{i}]\rangle$ cannot be generally correct and does not necessarily measure dissipation. In anticipation that waiting times do \emph{not} contribute to dissipation, we ignore them in the inference of $\langle\dot{S}_k[\hat{x}_i]\rangle$. Later, we also prove that they indeed result in a spurious contribution to the estimate for $\langle\dot{S}[x]\rangle$. 

Taking a single \emph{ergodically long}
trajectory, we use Eq.~\eqref{Ssemi1} or Eq.~\eqref{Ssemi} for the
microscopic, lumped, and post-lump milestoned trajectories to estimate the entropy production. To determine the accuracy of the estimates, we calculate the quality factor $0\leq Q \leq 1$, equal to the ratio of the estimated entropy production to the true entropy production of the microscopic process:
\begin{figure*}[t!]
    \centering
    \includegraphics[width=0.85\textwidth]{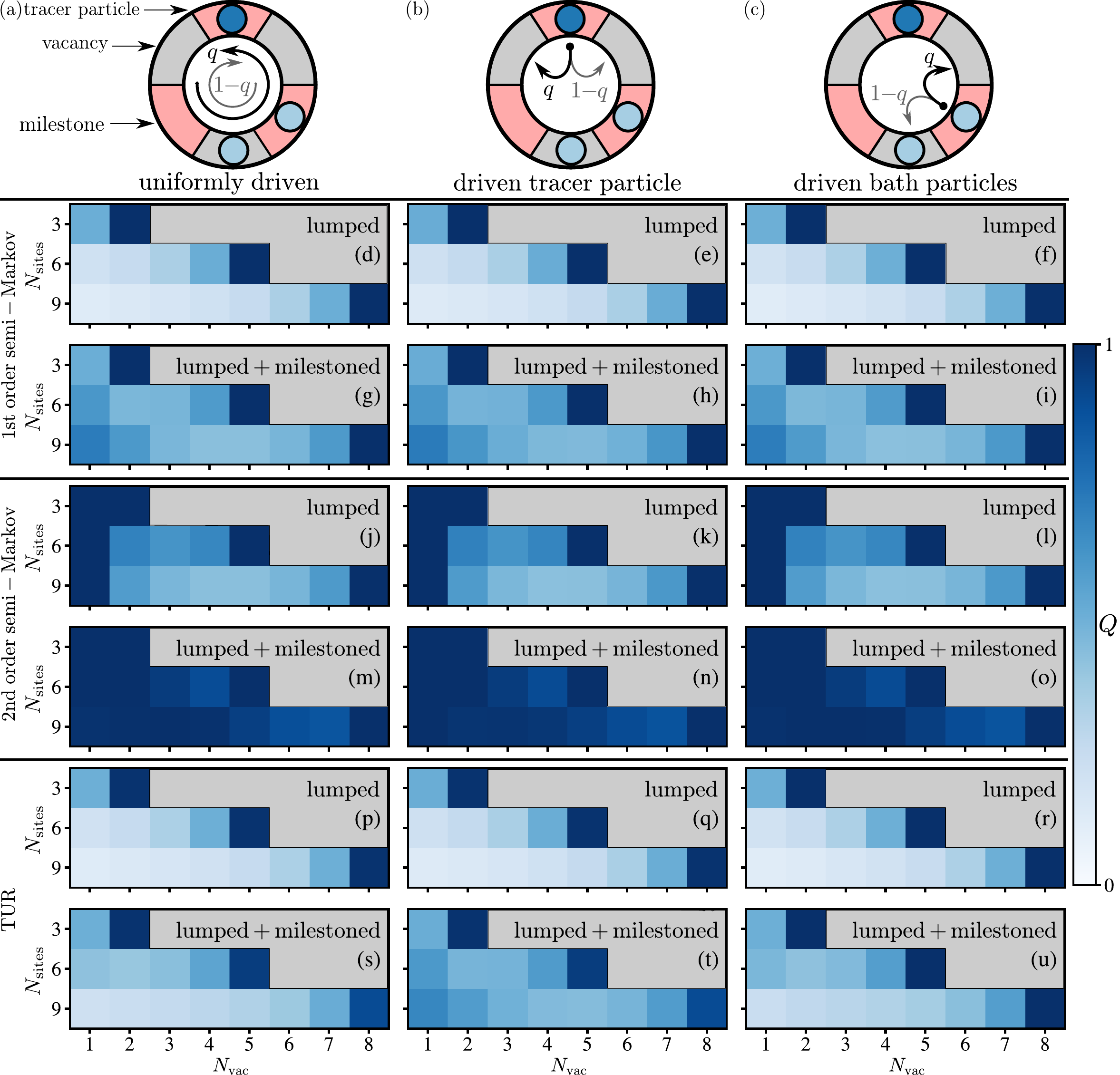}
    \caption{\textbf{Estimating the entropy production in single-file
        diffusion for different driving schemes.} (a-c) Schematic of
      the different driving scenarios, where the tracer particle is
      indicated in dark blue, and bath particles in light blue. Red
      sites indicate the milestone positions for the tracer
      particle. Asymmetric transition rates are
      $\{1{-}q,q\}{=}\{0.55,0.45\}$. In the schematics $N_{\rm
        sites}{=}6$ and $N_{\rm vac}{=}3$. (d-i) Quality factor of the
      entropy production estimated with the $1^{\rm st}$ order semi-Markov
      approximation (see Eq.~\eqref{Ssemi1}). (j-o) Quality factor of the entropy production
      estimated with the $2^{\rm nd}$ order semi-Markov approximation
      (see Eq.~\eqref{Ssemi}). (p-u) Quality factor of the entropy production estimated with the discrete-time TUR given by Eq.~\eqref{Qtur} using the tracer particle current.}
    \label{singlefile}
\end{figure*}
\begin{equation}
    Q = \langle \dot{S}^{\rm aff}_k[\hat{x}_i] \rangle/\langle \dot{S}[x] \rangle,
    \label{Q}
\end{equation}
where $i=\{ \rm lump, mil\}$.
We apply the $1^{\rm st}$ order (i.e., $k=1$) semi-Markov and $2^{\rm nd}$  order (i.e., $k=2$) semi-Markov
approximation to estimate $\langle \dot{S}[\hat{x}_{i}]\rangle $,
and the results are shown in Fig.~\ref{ringresults}. For $N_{\rm hid}\ge 3$ the estimates from
both coarse-grained trajectories (i.e.,~lumped and milestoned) are significantly lower than the true
entropy-production rate. This is due to the presence of hidden dissipative
cycles. 
Nevertheless, within the $1^{\rm st}$ order semi-Markov approximation
milestoning the lumped trajectory improves the entropy-production
estimate (compare Figs.~\ref{ringresults}a,b). On the $2^{\rm
  nd}$ order semi-Markov level milestoning does not improve the
estimate (Figs.~\ref{ringresults}c,d), since %$\hat{x}_{i}(t)$
the sequence of lumps (milestones, respectively)
is exactly a $2^{\rm nd}$ order semi-Markov chain. 

When there are no hidden cycles, i.e., for
$N_{\rm hid}=2$, we recover the exact microscopic entropy production with
Eq.~\eqref{Ssemi} which, notably, does \emph{not} take into account the
waiting-time contribution, i.e. $\langle \dot{S}^{\rm aff}_{2}[\hat{x}_{i}]\rangle=\langle \dot{S}[x] \rangle$. In fact, since
$\langle \dot{S}^{\rm wt}_2[\hat{x}_{\rm mil}]\rangle\ge 0$ for asymmetrically placed milestones, including the
waiting-time contribution from 
\cite{martinez2019inferring} violates the condition \eqref{ineq} and
may erroneously yield a positive entropy
production even when the microscopic dynamics obeys detailed balance (see
SM). This underscores that Eq.~\eqref{S} with $\{\theta \hat{x}(\tau)\}_{0\le
    \tau\le t}{\overset{!}{=}}\{\hat{x}(t-\tau)\}_{0\le \tau\le t}$ \emph{a priori} does \emph{not}
measure violation of detailed balance of processes with memory.

Interestingly, milestoning lumped dynamics may improve the entropy-production
estimate even in the presence of hidden cycles if the range of memory
is not captured exactly in the evaluation of Eq.~\eqref{S}. Notably, in
practice the extent of memory is unlikely to be known, and
computations at orders higher than 2 quickly become unfeasible.

%\textit{Driven single-file diffusion.---}
\section*{Driven single-file diffusion}
  Next, we consider
tracer-particle dynamics in discrete-time single-file diffusion %driven
on a ring (see Fig.~\ref{singlefile}), a paradigmatic model of dynamics with long memory and
anomalous diffusion
\cite{harris1965diffusion,lebowitz1967kinetic,Eli,Eli_2,Lapolla_2018,lapolla2020single,TUR_David} as well as a paradigm for
diffusion in crowded media
\cite{Olivier,Illien,Bertrand,shin2020biased,sokolov2012models} and
transport through biological channels \cite{hummer2001water}. Let
$N_{\rm sites}$ be the total number of sites on the ring, and $N_{\rm
  vac}$ the number of empty sites (vacancies), such that the total number of
particles is $N=N_{\rm sites}-N_{\rm vac}$.  At each discrete time step
a particle is picked at random and shifted with a given
probability to the left or right if the new site is unoccupied. Note that the latter condition introduces waiting times between jumps, i.e.,\ it effectively gives rise to local Poissonian clocks in the limit of continuous time. We
track the position of a tracer particle (see
e.g.,~Fig.~\ref{singlefile}a, dark blue), which results in a lumped
process, where multiple microscopic states correspond to the same location of the tracer particle but different arrangements of the unobserved ``bath'' particles.

We consider three different driving scenarios: (i) a uniformly driven
system with clockwise ($w_{+}$) and counterclockwise ($w_{-}$)
transition rates $\{w_{+},w_{-}\}{=}\{1{-}q,q\}$
(Fig.~\ref{singlefile}a), (ii) a driven tracer particle with
transition rates $\{w^{\rm t}_{+},w_{-}^{\rm t}\}{=}\{1{-}q,q\}$ and
bath particles with symmetric hopping rates
$\{w_{+},w_{-}\}{=}\{1/2,1/2\}$ (Fig.~\ref{singlefile}b) \cite{PhysRevE.54.3165, Lobaskin_2022, Miron_2021, Miron_2020, Kundu_2016}, and (iii)
the opposite scenario where the bath particles are driven, and the tracer
particle hops symmetrically (Fig.~\ref{singlefile}c) \cite{Banerjee_2022}. 

We further consider 3 maximally separated post-lumped milestones on
the ring (see e.g.,~Fig.~\ref{singlefile}a, red sites). This type of
milestoning corresponds to that implemented in
\cite{hartich2021violation} and yields
the exact entropy production in the case of a single particle
\cite{PhysRevX.11.041047,hartich2021violation}. 

In each driving scenario we evaluate, numerically, the exact
(microscopic) entropy production $\langle
\dot{S}[x] \rangle$ and compare it with $\langle \dot{S}^{\rm aff}_k[\hat{x}_i]
\rangle$ assuming $1^{\rm st}$ order (i.e.,\ $k=1$) (Fig.~\ref{singlefile}d-i)
and $2^{\rm nd}$ order (i.e.,\ $k=2$) semi-Markov (Fig.~\ref{singlefile}j-o) statistics for the
lumped and post-lumped milestoned trajectories, respectively. For all driving
scenarios, and both semi-Markov approximations, we find that
milestoning of the lumped trajectories significantly improves the entropy-production
estimate.  

For a further comparison, we also estimate  $\langle \dot{S}[\hat{x}_i]
\rangle$ via the \emph{discrete-time} TUR
\cite{Proesmans_2017}, which provides a universal trade-off between
precision of any thermodynamic flux and dissipation in
the system
\cite{PhysRevLett.114.158101,Proesmans_2017,PhysRevLett.130.087101}. By
determining the average $\langle J_{i} \rangle$ and variance ${\rm
  var}(J_{i})$ of the tracer-particle current in the lumped and post-lumped milestoned trajectories
(see SM for details), the
quality factor of the discrete-time TUR \cite{Proesmans_2017}
estimate reads
\cite{Proesmans_2017}  
\begin{equation}
       Q_{\rm TUR}=\ln{\left(2\langle J_{i} \rangle^{2}/{\rm var}(J_{i}) +1\right)}/\langle \dot{S}[x]\rangle,
    \label{Qtur}
\end{equation}
and is shown in Fig.~\ref{singlefile}p-u.  Note that for lumped dynamics it was proven that $0\leq Q_{\rm TUR} \leq 1$ \cite{Proesmans_2017}, but is a priori not clear if the TUR holds also for the post-lumped milestoned dynamics. Here we simply assume this to be true, and find that the %post-lumped 
milestoned estimates outperform the lumped estimates (see bottom panel in Fig.~\ref{singlefile}).
\begin{figure}[b!]
    \centering
     \includegraphics[width=8cm]{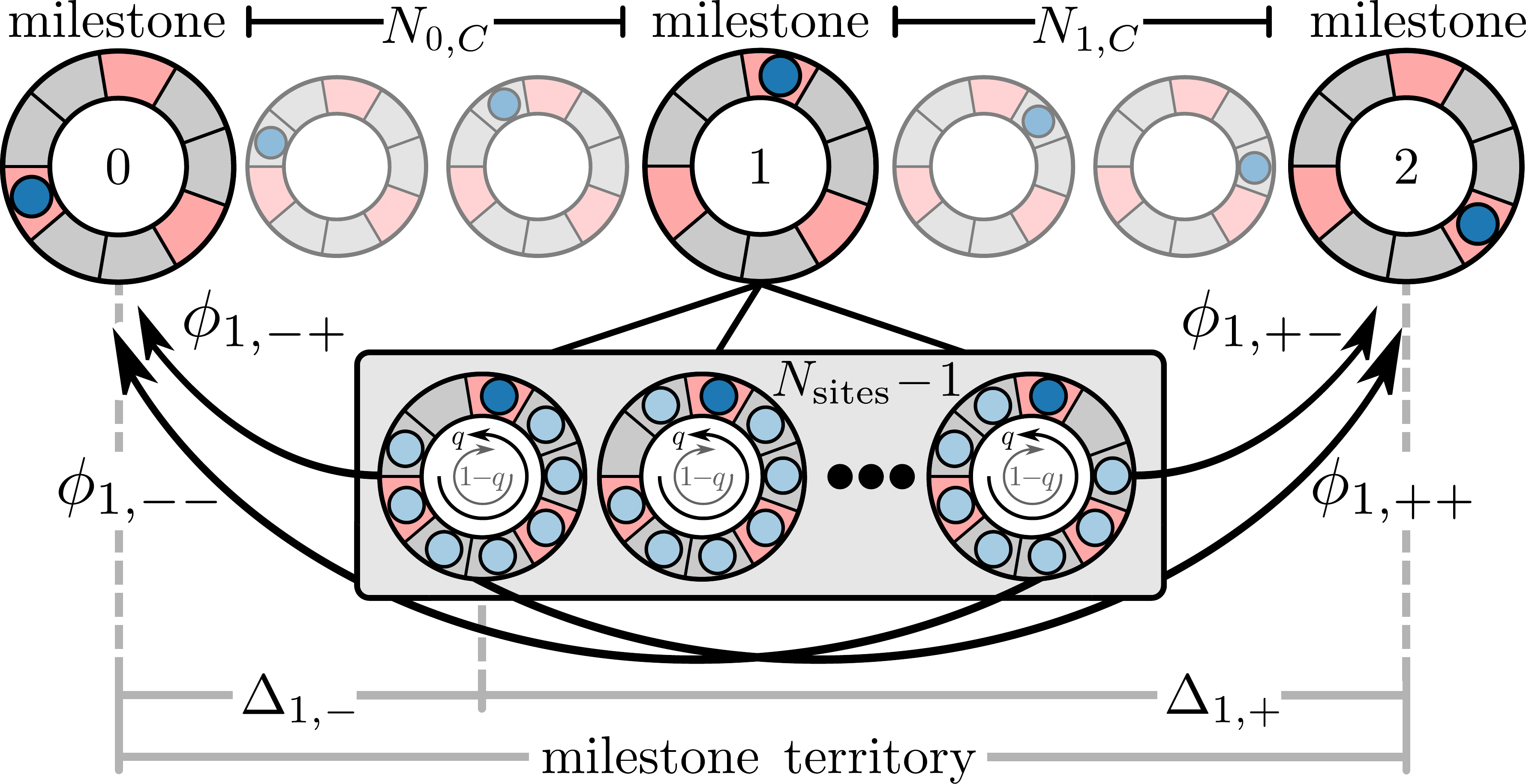}
    \caption{\textbf{Splitting probabilities in milestoned single-file
        diffusion.} The upper panel depicts the lumped states, where we
      only observe the location of the tracer particle. The lower
      panel depicts the $N_{\rm sites}{-}1$ hidden
      microstates. Adjacent milestones are separated by $N_{j,C}(N_{\rm
        sites}{-}1)$ microscopic states. Here we have $N_{0,C}=N_{1,C}=2$ and
      $N_{\rm sites}=9$. The random walk on milestones is an
      $2^{\rm nd}$ order semi-Markov process, with the sequence of
      milestones visited by the random walker determined by the 
      splitting probabilities $\phi_{j,\pm\pm}$.}
    \label{singlefile2}
\end{figure}

%\emph{Increasing ignorance.---}
\section*{Increasing ignorance}
In Fig.~\ref{ringresults}b we observe
that by increasing $N_{\rm lumps}$ at fixed $N_{\rm hid}$, which
increases the average distance between the milestones and hence our ``ignorance'', the $1^{\rm st}$ order estimate $\langle \dot{S}^{\rm aff}_1[\hat{x}_{\rm mil}]\rangle$ gradually improves. Similarly, in Fig.~\ref{singlefile}g-i
we find that for larger $N_{\rm sites}$ (except for $N_{\rm
  sites}{-}N_{\rm vac}=1$) the quality factor of the $1^{\rm st}$ order semi-Markov approximation is improved for the milestoned trajectory. On the contrary, the entropy estimates for the lumped dynamics deteriorate with increasing $N_{\rm sites}$ (see Fig.~\ref{singlefile}d-f). We now show, at
first sight paradoxically, that when milestoning is used, a
larger distance between  milestones and hence ``more ignorance'' may
indeed improve thermodynamic inference. %if the range of memory
%is not captured exactly (i.e.\
% in Eq.~\eqref{S}. 

Let us focus on uniformly driven single-file diffusion with one vacancy, i.e.~$N_{\rm vac}=1$, as shown in Fig.~\ref{singlefile2}. For a given number of sites $N_{\rm sites}$,
there are $N_{\rm sites}{-}1$ microstates with a fixed location of the tracer particle. There are in total $N_{\rm sites}(N_{\rm sites}{-}1)$ microscopic states, with groups of $N_{\rm sites}{-}1$ states forming the lumped states of the macroscopic system (i.e.,~the position of the tracer particle). We now designate some of those coarse states to
be milestones. Specifically, suppose we take $3$ milestones with $N_{j,C}$ coarse
states between milestone \ $j \ ({\rm mod} \ 3)$ and milestone $j+1 \ ({\rm mod} \ 3)$ (see Fig.~\ref{singlefile2}). From here on we always take the index $j$ modulo $3$.  Note that $N_{j,C}$ and
$N_{\rm sites}$ are related by $\sum_{j}[1+N_{j,C}]=N_{\rm sites}$.

In the
microscopic description of the problem, one can enumerate the microstates such that each milestone itself is
composed of $N_{\rm sites}{-}1$ states aligned along a line. We apply 
milestoning to obtain the trajectory $\hat{x}_{\rm mil}(t)$
as described before: the position of the tracer particle corresponds
to the last visited milestone, resulting in a $2^{\rm nd}$ order semi-Markov process. Specifically, the milestone sequence can be described by the conditional splitting probabilities \emph{in the sequence of distinct states} (i.e.~upon removing repeated consecutive milestoned states) $\{\phi_{j,++},\phi_{j,+-},\phi_{j,-+},\phi_{j,--}\}$, where $\phi_{j,++}$, for example, is the conditional probability of making a step from milestone $j$ to milestone $j+1$, given that the previous step occurred from milestone $j-1$ to milestone $j$. 

To calculate the splitting probabilities $\phi_{j,++}$ and $\phi_{j,-+}$,
for example, we consider the random walker starting in the leftmost
state of the segment of $N_{\rm sites}{-}1$ forming a milestone (see
gray box in Fig.~\ref{singlefile2}). If we denote this state by $k$, 
then $\phi_{j,-+}$ and $\phi_{j,++}$ are the probabilities to exit the interval
$\{k-\Delta_{j,-},...,k+\Delta_{j,+}\}$ through its left and right
boundaries with   
$\{\Delta_{j,-},\Delta_{j,+}\}{=}\{N_{j-1,C}(N_{\rm
  sites}{-}1){+}1,(N_{j,C}{+}1)(N_{\rm sites}{-}1)\}$.
One may recognize
that the solution to this problem is equivalent to solving the
Gambler’s ruin problem \cite{vrbikgambler}. Specifically, if we let $\alpha_{q}\equiv q/(1-q)$, we obtain
\begin{align}
    \phi_{j,++}&=(1{-}\alpha_{q}^{\Delta_{j,-}})/(1{-}\alpha_{q}^{\Delta_{j}}) , \ \phi_{j,-+}=1{-}\phi_{j,++}, \nonumber \\
    \phi_{j,+-}&=(1{-}\alpha_{q}^{\Delta_{j-1,+}})/(1{-}\alpha_{q}^{\Delta_{j}}), \ \phi_{j,--}=1{-}\phi_{j,+-}.
    \label{phi++}
\end{align}
where $\Delta_{j}=\Delta_{j,-}{+}\Delta_{j,+}$ denotes the territory of milestone $j$, i.e.,\ the domain size the trajectory needs to exit upon entering another milestone. Given these splitting probabilities, we can determine the entropy production at the level of the $1^{\rm st}$ order semi-Markov approximation, which reads
\begin{equation}
    \!\!\!\langle \dot{S}^{\rm aff}_{1} \rangle {=} \frac{1}{\langle \tau \rangle}\!\sum_{j}[\pi_{j}p_{j,+}{-}\pi_{j+1}p_{j+1,-}]\ln{\!\left(\frac{\pi_{j}p_{j,+}}{\pi_{j+1}p_{j+1,-}}\right)},\!
    \label{S1}
\end{equation}
where $\pi_{j}$ is the steady-state probability in milestone $j$, $p_{j,\pm}$ the transition probability to jump from milestone $j$ to $j \pm 1$, and $\langle \tau \rangle{=}\sum_{j}\langle \tau_{j} \rangle$ the total average waiting time in the milestones. Note that for equidistant milestones $\pi_{j}=1/3$, and Eq.~\eqref{S1} reduces to Eq.~\eqref{Ssemi1}.

The steady-state and transition probabilities can be determined from the conditions
\begin{align}
    \pi_{j}&=\pi_{j+1}p_{j+1,-}+\pi_{j-1}p_{j-1,+}, \nonumber \\
    \pi_{j}p_{j,+} &= \pi_{j-1}p_{j-1,+}\phi_{j,++}+\pi_{j+1}p_{j+1,-}\phi_{j,+-},
    \label{piit}
\end{align}
together with $\sum_{j}\pi_{j}=1$ and $p_{j,+}{+}p_{j,-}=1$. Upon inserting Eq.~\eqref{phi++} into Eq.~\eqref{piit} we find 
\begin{align}
    \pi_{j}&=(1-\alpha^{\Delta_{j-1,-}}_{q})(1-\alpha^{\Delta_{j,-}+\Delta_{j+1,-}}_{q})/\mathcal{N}, \nonumber \\
    p_{j,+}&=(1-\alpha^{\Delta_{j,-}}_{q})/(1-\alpha^{\Delta_{j,-}+\Delta_{j+1,-}}_{q}),
    \label{sol}
\end{align}
where $\mathcal{N}=\sum_{j}(1-\alpha^{\Delta_{j-1,-}}_{q})(1-\alpha^{\Delta_{j,-}+\Delta_{j+1,-}}_{q})$.

The average waiting time in a milestone can be split in two parts, with respect to entering the milestone from the left or the right. This yields
\begin{equation}
    \langle \tau_{j} \rangle = \pi_{j-1} p_{j-1,+}\langle \tau^{-}_{j} \rangle+\pi_{j+1} p_{j+1,-}\langle \tau^{+}_{j} \rangle,
    \label{tau1}
\end{equation}  
where the average waiting times conditioned on
starting from the left/right entrance of the milestone are given by (see page 4 in \cite{vrbikgambler})
\begin{align}
    \langle \tau^{-}_{j} \rangle &= (\Delta_{j,-}-\Delta_{j}\phi_{j,++})N/(2q-1), \nonumber \\ \langle \tau^{+}_{j} \rangle &= (\Delta_{j-1,+}-\Delta_{j}\phi_{j,+-})N/(2q-1),
\end{align}
and lead to the total average waiting time in the milestones
\begin{equation}
    \langle \tau \rangle = \frac{N\sum_{j}\Delta_{j,+}}{(1-2q)(3-2\sum_{j}1/(1-\alpha^{-\Delta_{j},-}_{q}))}.
    \label{tau}
\end{equation}
Inserting Eqs.~\eqref{sol} and \eqref{tau} into Eq.~\eqref{S1} we finally obtain the affinity contribution of the milestoned process within the $1^{\rm st}$ order semi-Markov approximation
\begin{align}
    \langle \dot{S}^{\rm aff}_{1}[\hat{x}_{\rm mil}] \rangle &=
    \frac{3/(N_{\rm sites}-1)+\sum_{j}N_{j,C}}{3+\sum_{j}N_{j,C}}\langle \dot{S}[x] \rangle \nonumber \\
     &=
    \frac{3/(N_{\rm sites}-1)+N_{\rm sites}-3}{N_{\rm sites}}\langle \dot{S}[x] \rangle
    \label{Ssol}
\end{align}
where $\langle \dot{S}[x] \rangle$ is the exact entropy production which can be obtained from the dynamics of the vacancy, and reads 
\begin{equation}
    \langle \dot{S}[x] \rangle = \frac{2q-1}{N}\ln{\left(\frac{q}{1-q}\right)}.
\end{equation}
We immediately see that Eq.~\eqref{Ssol} yields the exact entropy production
in the limit $N_{\rm sites}\rightarrow\infty$. This behavior can be
understood, qualitatively, as follows: as the total inter-milestone spacing increases, the relative
difference between splitting probabilities such as, e.g., $\phi_{j,+-}$ and $\phi_{j,++}$ becomes negligible because the relative milestone ``size'' (in the microscopic view) is negligible compared to the milestone territory, which suppresses correlations between entry- and exit-directions to and from the milestone, respectively. In other words, the milestone trajectory
approaches a $1^{\rm st}$ order semi-Markov process. But in the latter case,
we expect milestoning to become exact
\cite{hartich2021violation}.

For the "bare" lumped trajectory we simply set $N_{j,C}=0$ in the first line of Eq.~\eqref{Ssol}, which gives 
\begin{equation}
    \langle \dot{S}^{\rm aff}_{1}[\hat{x}_{\rm lump}] \rangle =\frac{1}{N_{\rm sites}-1}\langle \dot{S}[x] \rangle.
\end{equation} 
Hence, in the limit $N_{\rm sites}\rightarrow \infty$ the $1^{\rm st}$ order semi-Markov entropy production estimate in the lumped dynamics vanishes, as observed in Fig.~\ref{singlefile}d-f. 
\begin{figure*}[t!]
    \centering
     \includegraphics[width=0.95\textwidth]{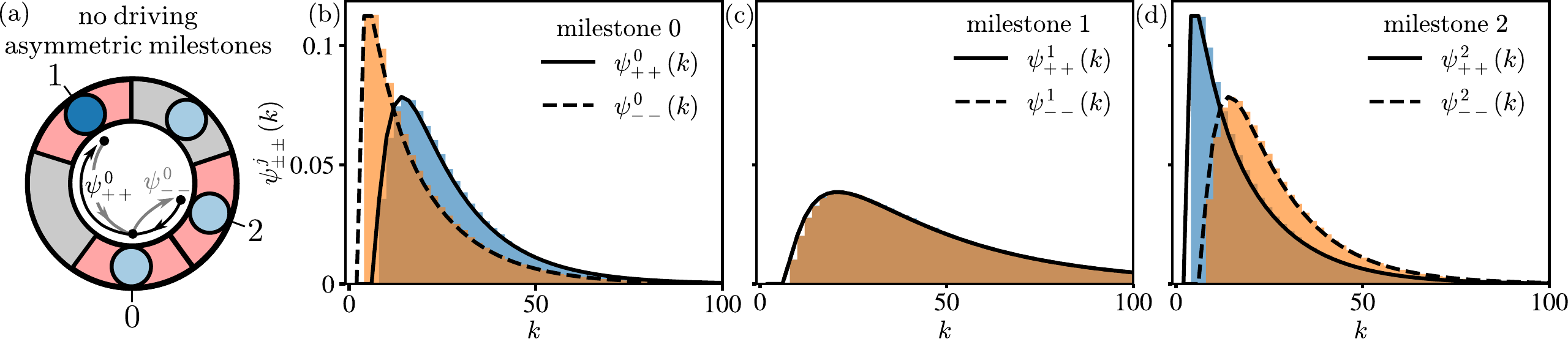}
    \caption{\textbf{Asymmetrically placed milestones result in asymmetric waiting-time distributions irrespective of the driving.} (a) Symmetric single file diffusion, that is, with no driving (i.e.,~$q=1/2$), for $N_{\rm sites}=5$, and $N_{\rm vac}=1$. The three milestones correspond to the positions of the tracer particle (dark blue) indicated in red. (b-d) Waiting-time distributions $\psi^{j}_{\pm\pm}(k)$ in each of the milestones obtained by simulations (blue + orange) and by Eq.~\eqref{psi} (full + dashed black lines). }
    \label{asymmetricsinglefile}
\end{figure*}

\section*{Spurious waiting-time contribution and kinetic hysteresis}
Notably, using Eq.~\eqref{Ssemi} on the uniformly driven single-file system with one vacancy, we recover the exact entropy production, i.e.,
\begin{equation}
     \langle \dot{S}^{\rm aff}_{2}[\hat{x}_{\rm mil}] \rangle = \langle \dot{S}^{\rm aff}_{2}[\hat{x}_{\rm lump}] \rangle = 
     \langle \dot{S}[x] \rangle,
\label{all}     
\end{equation}
which we also observe in Fig.~\ref{singlefile}j-o at $N_{\rm vac}=1$. 

Importantly, in Eq.~\eqref{Ssemi}  we have systematically discarded the waiting-time
contribution $ \langle \dot{S}^{\rm wt}_{2}[\hat{x}_{i}] \rangle $ that follows from naively applying
Eq.~\eqref{S} to non-Markovian trajectories
\cite{martinez2019inferring}. A direct computation shows \cite{martinez2019inferring}
\begin{equation}
\langle \dot{S}^{\rm wt}_{2}[\hat{x}_{\rm mil}] \rangle =\frac{1}{\langle \tau \rangle} \sum_{j}\sum_{\pm}p_{j,\pm\pm}D[\psi^{j}_{\pm\pm}||\psi^{j}_{\mp\mp}]\ge 0,
\label{wait_cont}
\end{equation}
where $p_{j,\pm\pm}{=}\pi_{j}p_{j,\pm}\phi_{j,\pm\pm}$, and the Kullback-Leibler divergence is given by
\begin{equation}
    D[\psi^{j}_{\pm\pm}||\psi^{j}_{\mp\mp}]\equiv\sum_{k\ge
  1}\psi^{j}_{\pm\pm}(k)\ln[\psi^{j}_{\pm\pm}(k)/\psi^{j}_{\mp\mp}(k)],
\end{equation} 
which is taken between the conditional
waiting-time distributions for consecutive forward $\psi^{j}_{++}(k)$ and backward
$\psi^{j}_{--}(k)$ transitions between
milestones. Note that as soon as the conditional forward and backward waiting time distributions are distinct ("asymmetric"), i.e.\ $\psi^{j}_{++}(k)\ne\psi^{j}_{--}(k)$ for some $k$, Eq.~\eqref{wait_cont} implies dissipation, irrespective of whether the microscopic dynamics is truly dissipative or not. The conditional waiting-time distributions read \cite{feller1991introduction} 
\begin{align}
  \psi^{j}_{++}(k)&=\mathcal{F}_{\Delta_{j}}(q,\Delta_{j,+},k)/\phi_{j,++}, \nonumber \\
%  \psi^{j}_{+-}(k)&=\mathcal{F}_{\Delta_{j}}(q,\Delta_{j+1,-},k)/\phi_{j,+-}, \nonumber \\
  \psi^{j}_{--}(k)&=\mathcal{F}_{\Delta_{j}}(1-q,\Delta_{j-1,+},k)/\phi_{j,--}, \nonumber %\\
%  \psi^{j}_{-+}(k)&=\mathcal{F}_{\Delta_{j}}(1-q,\Delta_{j,-},k)/\phi_{j,-+},
  \label{psi}
\end{align}
where we introduced the auxiliary function
\begin{align}
\!\!\mathcal{F}_\Delta(q,z,k)&\equiv\frac{2^k}{\Delta}(1-q)^{\frac{k+z}{2}}q^{\frac{k-z}{2}}\nonumber
\\&\times
\sum^{\Delta-1}_{l=1}\cos^{k-1}\!\left(\frac{l\pi}{\Delta}\right)\sin\!\left(\frac{l\pi}{\Delta}\right)\sin\!\left(\frac{l\pi
  z}{\Delta}\right)\!.
\end{align}
In Fig.~\ref{asymmetricsinglefile} we show that as soon as the distance to the left/right neighboring milestone is not symmetric, we obtain ${D[\psi^{j}_{\pm\pm}||\psi^{j}_{\mp\mp}]>0}$ and thus it follows
from Eq.~\eqref{all} that $\langle \dot{S}^{\rm aff}_{2}[\hat{x}_{i}]\rangle +\langle
\dot{S}^{\rm wt}_{2}[\hat{x}_i]\rangle> \langle \dot{S}[x] \rangle$. Moreover, even under
perfect (microscopic) detailed balance (i.e.,\ $q{=}1/2$) we find
that the waiting-time distributions are asymmetric (see Fig.~\ref{asymmetricsinglefile}), and therefore $D[\psi^{j}_{\pm\pm}||\psi^{j}_{\mp\mp}]{>}0$ and thus
$\langle
\dot{S}^{\rm wt}_{2}[\hat{x}_{\rm mil}]\rangle{>}\langle \dot{S}[x] \rangle{=}0$.
%%delivers another---this time a $2^{\rm nd}$ order semi-Markov---example demonstrating that $\langle
%\dot{S}^{\rm wt}_{2}[\hat{x}_{\rm mil}]\rangle$ does not generally contribute to dissipation. 
%to the main conclusion in Ref.~\cite{martinez2019inferring}, which was meanwhile further used in \cite{nitzan2022universal} to construct a (what was declared as a) "lower" bound on the total entropy production. 
The above $2^{\rm nd}$ order semi-Markov example and the family of $1^{\rm st}$ order semi-Markov counterexamples in \cite{comment,PhysRevX.11.041047,Qian} disprove that waiting-times \emph{in general} contribute to dissipation in Eq.~\eqref{S}. The waiting-time contribution in Eq.~\eqref{wait_cont} in our example quantifies the difference between the waiting-time distributions between forward versus backward transitions, but it does \emph{not} contribute to dissipation.

Note that the post-lumped milestoned process displays kinetic hysteresis
(i.e.,\ coarse-graining and time reversal do not commute;
see SM)
\cite{PhysRevX.11.041047}, whereas the original lumped dynamics does
\emph{not}. For the latter, the waiting-time distributions are symmetric irrespective of the lump sizes and presence or absence of driving, resulting in a vanishing waiting-time contribution $\langle
\dot{S}^{\rm wt}_{2}[\hat{x}_{\rm lump}] \rangle =0$. Therefore, in the presence of kinetic
hysteresis in coarse-grained dynamics, the waiting-time contribution in Eq.~\eqref{wait_cont}
does \emph{not} contribute to dissipation, and in its absence it typically (i.e.\ in our examples) seems to vanish (see, however, non-vanishing contributions in \cite{martinez2019inferring}).~Therefore, since it is \emph{not} possible in practice to test for the presence of kinetic hysteresis without the knowledge of the full dynamics, asymmetric waiting-time distributions generally \emph{cannot} be used to infer dissipation via Eq.~\eqref{wait_cont}, i.e.,\ there is no implication between the two: $\langle\dot{S}[x]\rangle>0\centernot\implies\langle \dot{S}^{\rm wt}[\hat{x}] \rangle>0$ and in turn $\langle \dot{S}^{\rm wt}[\hat{x}] \rangle>0 \centernot\implies \langle\dot{S}[x]\rangle>0$. In fact, using the naive Markovian
time-reversal operation $\{\theta \hat{x}_i(\tau)\}_{0\le
    \tau\le t} {\overset{!}{=}} \{\hat{x}_i(t-\tau)\}_{0\le \tau\le t}$ in
Eq.~\eqref{S}, at least for $1^{\rm st}$ and $2^{\rm nd}$ order semi-Markov
processes in the presence of kinetic hysteresis, is inconsistent and does \emph{not}  describe
dissipation correctly. Notably, the complete thermodynamically consistent definition of dissipation for $1^{\rm st}$ order semi-Markov
processes has been established in \cite{PhysRevX.11.041047}, while for $2^{\rm nd}$ order semi-Markov
processes the affinity contribution $\langle
\dot{S}^{\rm aff}_{2}[\hat{x}] \rangle$ seems to provide a correct lower bound, $\langle
\dot{S}[x] \rangle\ge \langle
\dot{S}^{\rm aff}_{2}[\hat{x}] \rangle$.  

%\emph{Discussion.---}
\section*{Discussion}
\emph{Correct notion of time reversal.---} Contrasting the
setting where all hidden degrees of freedom relax instantaneously fast,
that is solved essentially completely within the notion of
``local detailed balance'' \cite{Massi}, inferring dissipation %violations of detailed balance
from general coarse observations
%\emph{in absence} of a time-scale separation
remains both conceptually and technically
challenging and is still  understood very poorly. Sufficiently far from
equilibrium, challenges appear even in the presence of a time-scale
separation between the hidden and observed degrees of freedom \cite{hartich2021violation}. In the absence of such time-scale
separation, and for an arbitrary extent of memory, the problem remains
unsolved. This is because a correct general mathematical definition of
dissipation in the presence of memory is not known. In particular,
the definition in Eq.~\eqref{S} is incomplete, because the correct general
form of the time-reversal operation $\theta$ remains elusive. Memory
reflects (anti-)persistence in dynamics, and it is therefore not
surprising that it manifests some abstract analogy to momenta. 

Whereas practical methods are available to infer the presence \cite{Dima_Test,Lindner} and
  range \cite{Toolbox, PhysRevResearch.5.L012026, 10.1063/5.0158930} of memory in coarse observations, the precise
 flavor of non-Markovianity is typically not known. In particular, if
 only coarse-grained trajectories are accessible, such as in
 experimental studies, it is inherently \emph{impossible} to check for
 the presence of kinetic hysteresis. Therefore, referring to the
 absence of kinetic hysteresis as a "mild assumption"
 \cite{Horrorwitz_mild} seems somewhat inappropriate. Either way,
 since $\langle \dot{S}^{\rm wt}[\hat{x}] \rangle$ can be positive
 even under detailed balance, and one cannot check for kinetic hysteresis, a non-vanishing $\langle \dot{S}^{\rm wt}[\hat{x}] \rangle$ generally cannot be used to infer dissipation.
 
 Even if the order of memory is %and the presence/absence of kinetic hysteresis were
 known, the correct notion of physical time reversal generally remains
 elusive. For semi-Markov
 processes of $1^{\rm st}$ order it has been proven that
 the transition-path periods (i.e.,\ the duration of successful transitions
 between coarse states) are "odd" under time reversal (i.e.,\ transition paths between any pair of states must be reverted in the time-reversed trajectory like momenta in inertial systems), and the
 thermodynamically consistent time-reversal operation was established
 \cite{PhysRevX.11.041047}. Importantly, by explicit counterexamples it was proven that waiting-time effects in $1^{\rm st}$ and $2^{\rm nd}$ order semi-Markov processes generally do
 \emph{not} measure dissipation. Note that reverting all transition paths in the time-reversed trajectories ensures $\langle\dot{S}^{\rm wt}[\hat{x}]\rangle=0$, and thus has the same effects as simply ignoring $\langle\dot{S}^{\rm wt}[\hat{x}]\rangle$. Moreover, we found that for the $2^{\rm nd}$ order semi-Markov process with no hidden dissipative cycles \footnote{A dissipative cycle in a Markov process is a cycle for which the Kolmogorov cycle criterion is violated. The cycle is said to be hidden if the states of said cycle are coarse-grained into the same lump.} in Fig.~\ref{singlefile} (see also
 \cite{hartich2021violation,comment}), the two-step affinity exactly recovers the microscopic dissipation, which requires all waiting time contributions to be systematically discarded.
 To what extent and under which conditions this, i.e., $\langle\dot{S}^{\rm aff}[\hat{x}_i]\rangle=\langle\dot{S}[x]\rangle$ in the absence of hidden cycles, is
 true for general $k^{\rm th}$ order semi-Markov
 processes ($k>2$) still needs to be established. Whether a general form of time reversal $\theta$ exists that would always yield a consistent waiting-time contribution to dissipation remains unknown. One can construct $2^{\rm nd}$ (but not $1^{\rm st}$) order semi-Markov processes without kinetic hysteresis that have asymmetric waiting time distributions whose contribution does not %necessarily 
 violate the inequality~\eqref{ineq} \cite{martinez2019inferring}. However, our results
 show that for $1^{\rm st}$  and $2^{\rm nd}$ order semi-Markov processes 
 waiting-time effects as encoded in $\langle\dot{S}^{\rm wt}[\hat{x}]\rangle$ are either nominally zero or else must be discarded, since one cannot know (nor can one test) for the presence of kinetic hysteresis without the knowledge of microscopic dynamics. We believe this to be a result of an incorrect or incomplete definition of time-reversal $\theta$ when applying Eq.~\eqref{S} to $1^{\rm st}$ and $2^{\rm nd}$ order semi-Markov processes.
We expect this to also be true  for processes
with longer memory, where, however, evaluating the path probabilities becomes challenging.     

\emph{Practical aspects of milestoning.---} From a practical perspective, one therefore does not generally know how
to correctly apply Eq.~\eqref{S} to infer dissipation from observed non-Markovian
trajectories. A convenient, practical, and consistent approach we push forward here is to
first milestone trajectories and thereafter assume
$1^{\rm st}$  or $2^{\rm nd}$ order semi-Markov correlations in the
evaluation of Eq.~\eqref{S} excluding any waiting-time
contributions. Even if this assumption is not satisfied exactly
(i.e.,\ the memory is actually longer-ranged), a \emph{meaningful} milestoning will
systematically render the observations 
%``more Markovian'' 
"closer" to the underlying microscopic dynamics
and thus improve thermodynamic
inference. Moreover, a sparser placement of milestones
increases the  accuracy of  inference and is numerically more efficient.

One is therefore tempted to think that milestones should generally be placed as sparsely
as possible for optimal thermodynamic inference. There is, however, a practical
limitation. Consider the example of observing a tracer particle in a
single-file. If we increase the spacing between milestones, the
estimate for $\langle \dot{S}[\hat{x}_{\rm mil}]\rangle$ per
milestone transition
indeed gradually improves (i.e.,\ increases). At large distances between milestones, however, the
sequence of visited milestones will appear nearly deterministic,
whereby transitions against the driving (i.e.,\ ``time-reversed
trajectory segments'') will become increasingly unlikely. Thus, the
evaluation of Eq.~\eqref{S} using trajectory-derived path
probabilities will require better and better sampling
(see the SM for a quantitative statement). Analytical
estimates of transition probabilities (such as presented here) are of course not affected. In general, the placement of
milestones should be optimized as a trade-off between accuracy and
statistical feasibility.

Finally, we stress that the proposed post-lumped
milestoning based thermodynamic
inference requires at least one dissipative cycle to be
observable. If some dissipative cycles are hidden by the coarse-graining,  milestoning based estimation will yield a \emph{true} lower
bound on dissipation. However, if truly \emph{all} dissipative cycles become
hidden, the method will fail, and one should resort to alternative
``transition-based'' approaches developed recently \cite{Snippets,
  van2022thermodynamic,Polettini, deguenther2023fluctuating} to infer some lower bound on
dissipation.

\textbf{Acknowledgments.---}This project has received funding from the German Research Foundation (DFG) through the Heisenberg Program (grant GO 2762/4-1 to AG),
%European Research Council (ERC) under the European agreement No 101086182 to AG), 
from the Robert A. Welch Foundation (Grant No. F- 1514 to DEM), the National Science Foundation (Grant Nos. CHE 1955552 to DEM and IIS 2212048 to EV), from the Adobe Inc., and from the Alexander von Humboldt Foundation.

\textbf{Author contributions.---}A.G.\ and D.E.M.\ conceived the research and drafted the manuscript. K.B.\ wrote the computer program and performed the analytical and numerical calculations. K.S.\ and E.V.\ contributed to theory development and provided critical feedback on the manuscript. All authors contributed to the interpretation of the results, and the writing of the manuscript. 

\textbf{Competing interests.---} 
The authors declare that they have no competing interests. 

\textbf{Data availability.---} The raw data and source codes to reproduce the results  shown in this manuscript are publicly available at \cite{Blom2023}.
%- --------------------------------
%\nocite{*} % Show all Bib-entries

\let\oldaddcontentsline\addcontentsline% Store \addcontentsline
\renewcommand{\addcontentsline}[3]{}% Make \addcontentsline a no-op
\bibliographystyle{apsrev4-1.bst}
\bibliography{makarov.bib}
\let\addcontentsline\oldaddcontentsline% Restore \addcontentsline

\clearpage
\newpage
\onecolumngrid
\renewcommand{\thefigure}{S\arabic{figure}}
\renewcommand{\theequation}{S\arabic{equation}}
\setcounter{equation}{0}
\setcounter{figure}{0}
\setcounter{page}{1}
\setcounter{section}{0}

\begin{center}\textbf{Supplemental Material for:\\ Milestoning estimators of dissipation in systems
  observed at a coarse resolution: When ignorance is truly
  bliss}\\[0.2cm]  
\end{center}  
%---------------------------------
%---------------------------------
In this Supplementary Material (SM) we present details and further extensions of the results shown in the main manuscript. 
%---------------------------------
%---------------------------------

%---------------------------------
%---------------------------------
\subsection{Toy model for a higher-order semi-Markov process}
%---------------------------------
%---------------------------------
In Fig.~\ref{S1}a we extend the toy model for a $2^{\rm nd}$ order semi-Markov process introduced in the main manuscript. The extension includes an additional exit and entrance connection between the lumped states with different transition rates $\{w^{\dagger}_{{\rm lump},+},w^{\dagger}_{{\rm lump},-}\}$. In Fig.~\ref{S1}b-c we observe that milestoning the lumped trajectory significantly improves the entropy production estimate within the $1^{\rm st}$ order semi-Markov approximation. On the $2^{\rm nd}$ order semi-Markov level, milestoning does not improve nor deteriorate the entropy-production estimate. 
\begin{figure}[h!]
    \centering
    \includegraphics[width=0.95\textwidth]{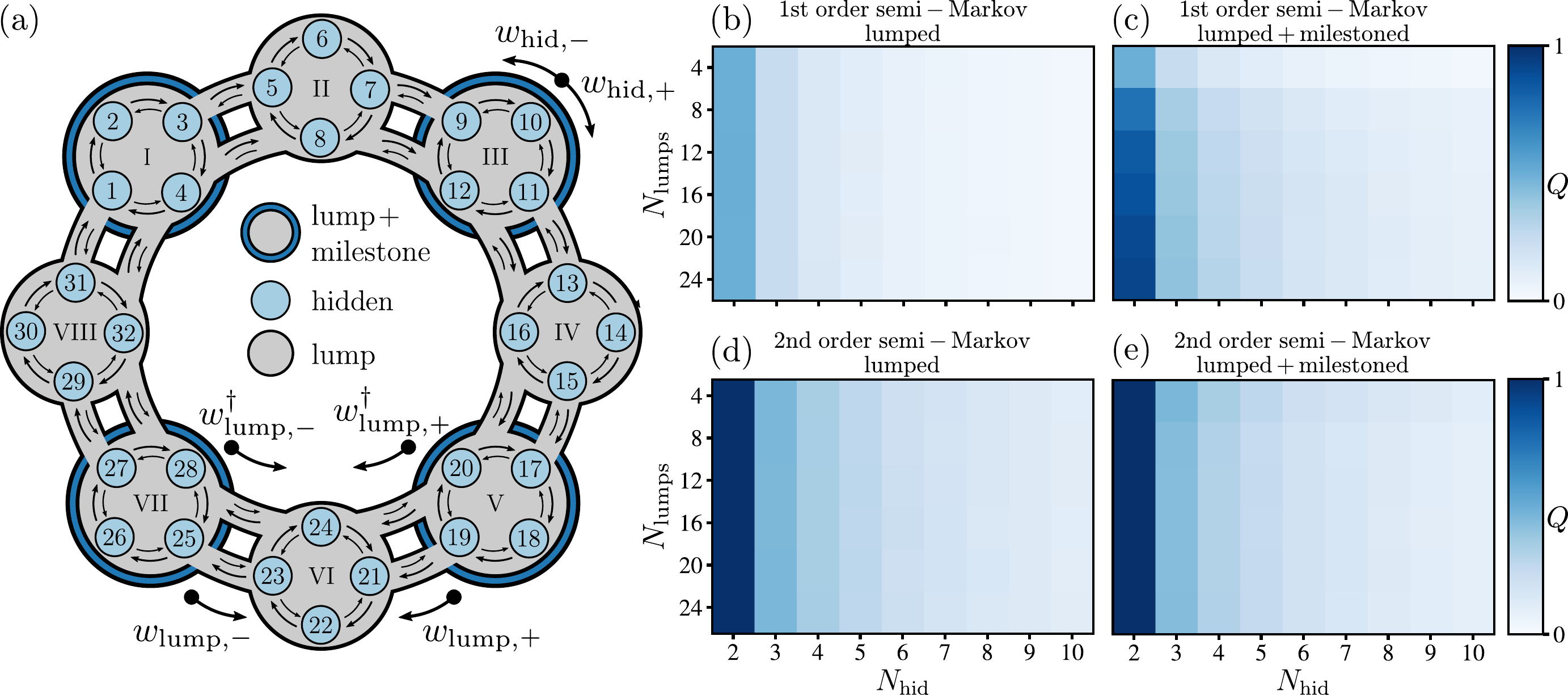}
    \caption{\textbf{Entropy production inference for a semi-Markov process beyond $2^{\rm nd}$ order} (a) Schematic of the toy model with 32
      microstates lumped into domains I-VIII
      (gray) each containing 4 microstates. Milestoning involves further post-processing of the lumped process: some
      lumped states (here, the domains I, III, V, and VII)
      are declared to be milestones (blue). Once the trajectory enters a
      milestone, the coarse-grained state of the  milestone
      trajectory remains in said milestone until it enters a different milestone. (b-e) Estimating the entropy production from a
        coarse-grained trajectory governed by the discrete-state ring-like Markov
      chain shown in (a), with $N_{\rm lumps}$ the
      number of lumped states, $N_{\rm hid}$ the number of hidden
      microstates per lumped state, transition rates
      $\{w_{{\rm lump},+},w_{{\rm lump},-}\}=\{w_{{\rm hid},+},w_{{\rm hid},-}\}=\{0.52,0.48\}$, $\{w^{\dagger}_{{\rm lump},+},w^{\dagger}_{{\rm lump},-}\}=\{0.51,0.49\}$, and $10^{8}$ discrete steps. (b, d) Quality factor $Q$ (see Eq.~(5) in manuscript) for the entropy production estimated with the
      lumped trajectory assuming $1^{\rm st}$ order semi-Markov (see Eq.~(3) in manuscript) (b) and $2^{\rm nd}$ order semi-Markov (see Eq.~(4) in manuscript) (d). (c, e) Quality factor for the entropy production
      estimated with the post-lumped milestoned trajectory with 4 equidistant milestones, assuming $1^{\rm st}$
      order semi-Markov (c) and $2^{\rm nd}$ order semi-Markov (e).
     }
    \label{S1}
\end{figure}
%---------------------------------
%---------------------------------
\subsection{Waiting-time contribution in the ring process}
\begin{figure}[b!]
    \centering
    \includegraphics[width=0.95\textwidth]{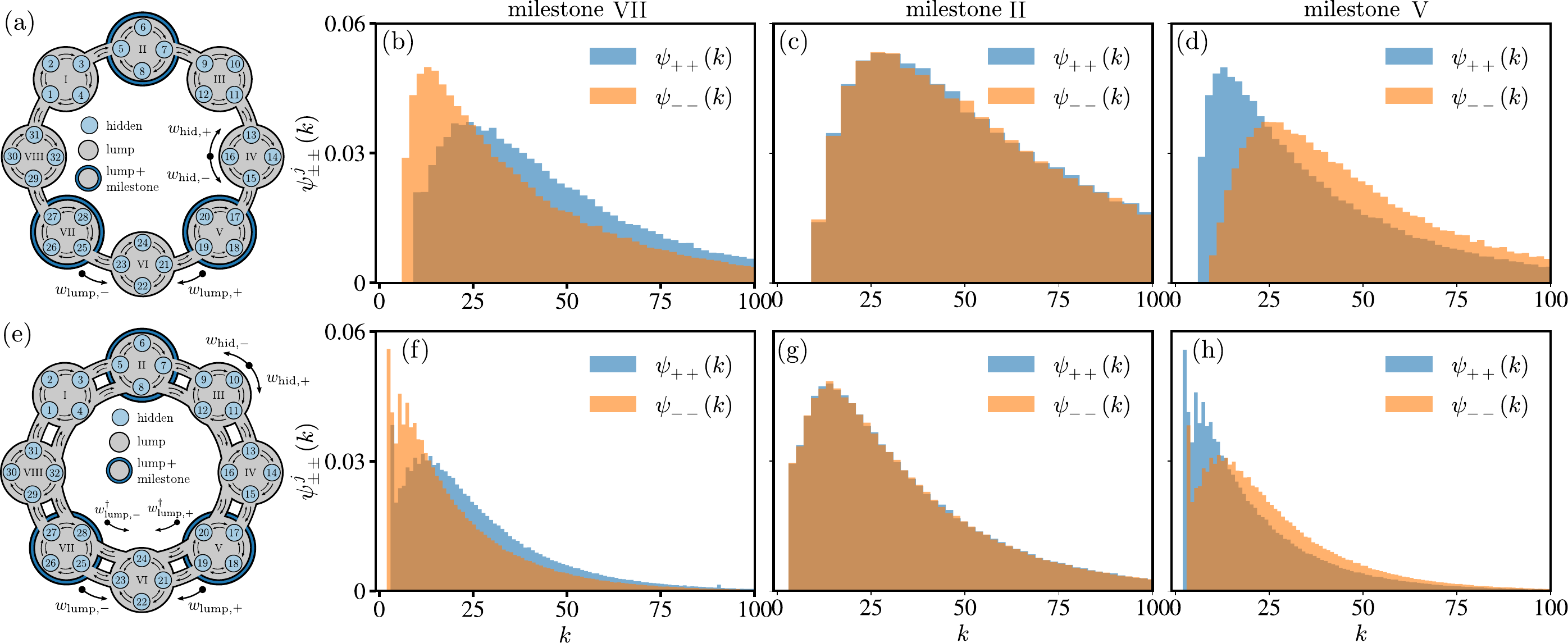}
    \caption{\textbf{Waiting-time distribution for $3$ asymmetrically placed  milestones in the ring process with detailed-balance.} In all panels, we consider symmetric transition rates (inside the lumps and those connecting the different lumps) $w_{{\rm hid},+}=w_{{\rm lump},+}=w^{\dagger}_{{\rm lump},+}=w_{{\rm hid},-}=w_{{\rm lump},-}=w^{\dagger}_{{\rm lump},-}=1/2$ for the ring process with $N_{\rm lump}=8$, $N_{\rm hid}=4$, and milestones placed at lumps $\{\rm II, V, VII\}$. Waiting-time distributions are obtained from a stochastic trajectory with $10^{8}$ discrete steps. (a-d) Waiting-time distributions for the toy model introduced in Fig.~1a in the main manuscript. (e-h) Waiting-time distributions for the toy model introduced in Fig.~\ref{S1}a in the SM.}
    \label{S2}
\end{figure}

%---------------------------------
%---------------------------------
In Fig.~\ref{S2} we show the waiting-time distributions for the toy models shown in Fig.~1a in the main manuscript and  Fig.~\ref{S1}a in the SM. For both models, we take $3$ milestones which are placed asymmetrically, i.e.~the milestones are \emph{not} placed equidistantly as shown in Fig.~\ref{S2}a,e. Furthermore, we take fully symmetric transition rates $w_{{\rm hid},+}=w_{{\rm lump},+}=w^{\dagger}_{{\rm lump},+}=w_{{\rm hid},-}=w_{{\rm lump},-}=w^{\dagger}_{{\rm lump},-}=1/2$, which renders both, the exact microscopic result as well as the affinity entropy estimate zero, i.e.~$\langle\dot{S}[x]\rangle=\langle \dot{S}^{\rm aff}_{2}[\hat{x}_{\rm lump}] \rangle=\langle \dot{S}^{\rm aff}_{2}[\hat{x}_{\rm mil}] \rangle = 0$. In Fig.~\ref{S2}b-d and f-h we observe that the waiting-time distributions $\psi_{++}(k)$ and $\psi_{--}(k)$ for milestones VII and V are unequal (compare orange and blue histograms). This renders the waiting-time contribution to the entropy production nonzero, i.e., (note that the index $j$ sums over the $3$ milestones)
\begin{equation}
    \langle \dot{S}^{\rm wt}_{2}[\hat{x}_{\rm mil}] \rangle =\frac{1}{\langle \tau \rangle} \sum_{j}\sum_{\pm}p_{j,\pm\pm}D[\psi^{j}_{\pm\pm}||\psi^{j}_{\mp\mp}]\ge 0.
    \label{wait_cont}
\end{equation}
Therefore, by incorporating Eq.~\eqref{wait_cont} in the total estimate for the entropy production, one would mistakenly conclude that this system (which strictly obeys detailed balance) has dissipative fluxes, which is clearly not the case.
%---------------------------------
%---------------------------------
\subsection{Inference via the Thermodynamic Uncertainty Relation}
%---------------------------------
%---------------------------------
Fig.~3p-u in the main manuscript depicts the inference of the entropy production rate in a single-file system using the discrete-time thermodynamic uncertainty relation (TUR). The discrete-time TUR reads \cite{Proesmans_2017S}
\begin{equation}
       \langle \dot{S}_{\rm TUR} \rangle =\ln{\left(2\langle J_{i} \rangle^{2}/{\rm var}(J_{i}) +1\right)},
    \label{Stur}
\end{equation}
where $\langle J_{i} \rangle$ and ${\rm
  var}(J_{i})$ are the average and variance, respectively, of the tracer-particle current in the lumped and post-lumped milestoned trajectories. To determine the current in the lumped and post-lumped milestones trajectories $\hat{x}_{i}(t)$, we simply take the difference between two consecutive jumps, i.e.,
  \begin{equation}
      J_{i}(t) = \hat{x}_{i}(t)-\hat{x}_{i}(t-1),
  \end{equation}
  where the difference is taken modulo the number of sites ($i={\rm lump}$) or milestones ($i={\rm mil}$). The mean and variance of the current are then determined via
  \begin{equation}
      \langle J_{i} \rangle = \frac{1}{T}\sum_{t=1}^{T} J_{i}(t), \ {\rm
  var}(J_{i}) = \frac{1}{T}\sum_{t=1}^{T}(J_{i}-\langle J_{i} \rangle)^{2},
  \end{equation}
  where $T$ is the total number of steps in the trajectory. For the results shown in Fig.~3p-u in the main manuscript we have $T=10^{8}$.
%---------------------------------
%---------------------------------
\subsection{Milestoning and kinetic hysteresis}
%---------------------------------
%---------------------------------
\begin{figure}[t!]
    \centering
    \includegraphics[width=0.99\textwidth]{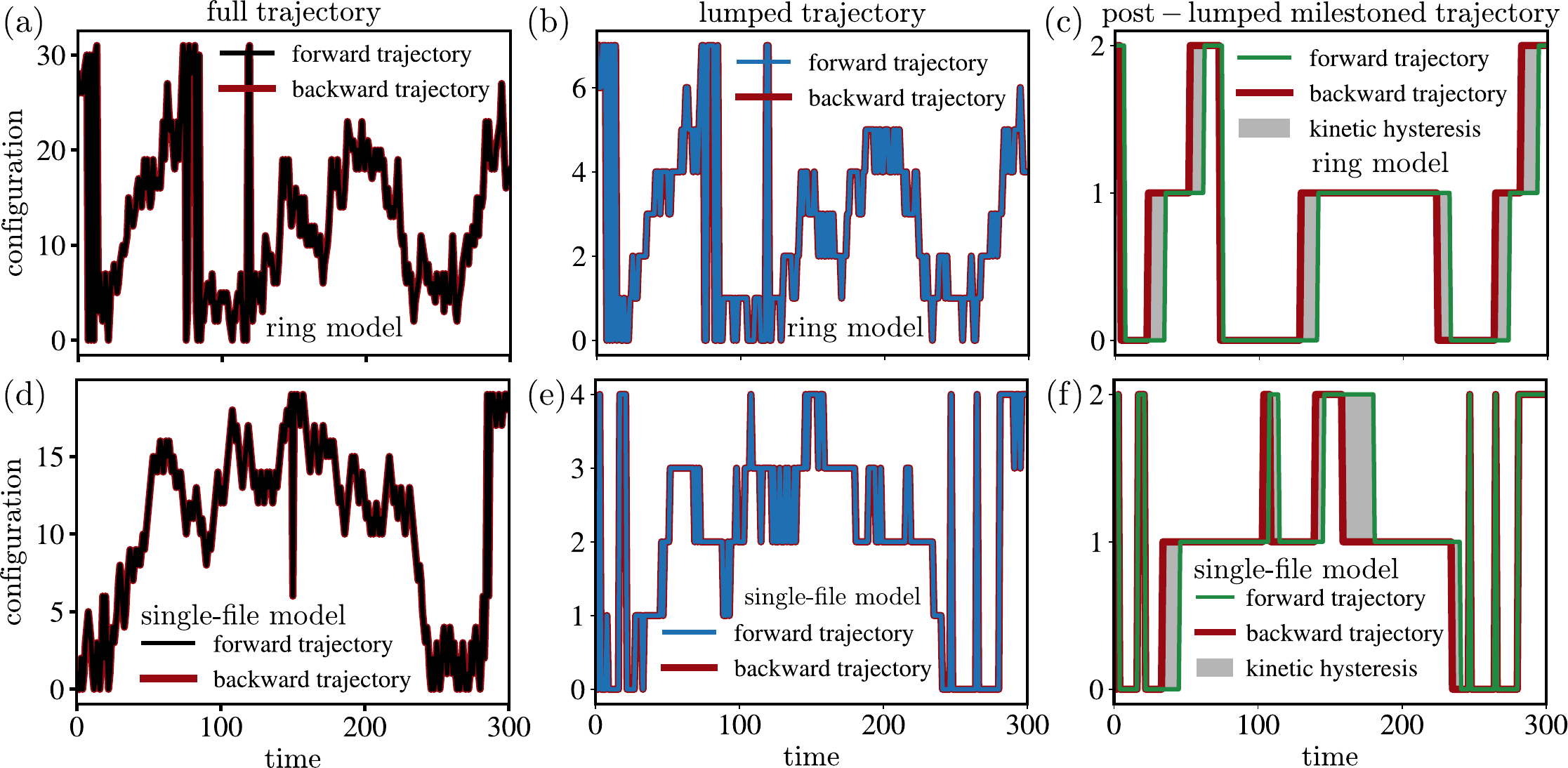}
    \caption{\textbf{Stochastic trajectories and kinetic hysteresis in the ring model and single-file system.} For the ring model, we consider $N_{\rm lump}=8$ lumped macrostates, and $N_{\rm hid}=4$ hidden microstates per lump. For the single-file system, we consider $N_{\rm sites}=5$ sites, and $N_{\rm vac}=1$ vacancies. Backward trajectories are indicated with the solid red line. (a, d) Microscopic stochastic trajectories. (b, e) Lumped stochastic trajectories. (c, f) Milestoned stochastic trajectories. The gray shaded area denotes the kinetic hysteresis where forward and backward trajectories differ. }
    \label{S3}
\end{figure}
A fundamental difference between lumping and milestoning is that for the latter, time reversal and coarse-graining do not commute, which was coined "kinetic hysteresis" \cite{PhysRevX.11.041047S}. To see this, let us consider a stochastic trajectory for the ring model (see Fig.~1a in the main manuscript) and single-file system (see Fig.~3 in the main manuscript), as shown in Fig.~\ref{S3}. In Fig.~\ref{S3}a,d we display the microscopic stochastic trajectories, for which the forward (black line) and backward (red line) trajectories are identical under time inversion. For the lumped trajectories, shown in Fig.~\ref{S3}b,e we also find that the backward trajectories traverse along identical paths under time inversion. However, it turns out that the milestoned trajectories depicted in Fig.~\ref{S3}c,f display the following phenomenon \cite{PhysRevX.11.041047S}: If we coarse grain the same trajectory backward in
time, we discover a kinetic hysteresis. That is, the time-reversed coarse trajectory (red line in Fig.~\ref{S3}c,f), where time is running from right to left, differs from the forward one (green line in Fig.~\ref{S3}c,f). Hence, using the naive Markovian time-reversal operation $\theta$
\begin{equation}
    \{\theta \hat{x}_i(\tau)\}_{0\le
    \tau\le t} \overset{!}{=} \{\hat{x}_i(t-\tau)\}_{0\le \tau\le t}\quad {\displaystyle \bot }
\end{equation} 
in the presence of kinetic hysteresis is inconsistent and does \emph{not} describe dissipation upon inserting into Eq.~(1) in the manuscript. 
%---------------------------------
%---------------------------------
\subsection{"Statistical burden" of improving thermodynamic inference via milestoning}
\begin{figure}[h!]
    \centering
    \includegraphics[width=0.99\textwidth]{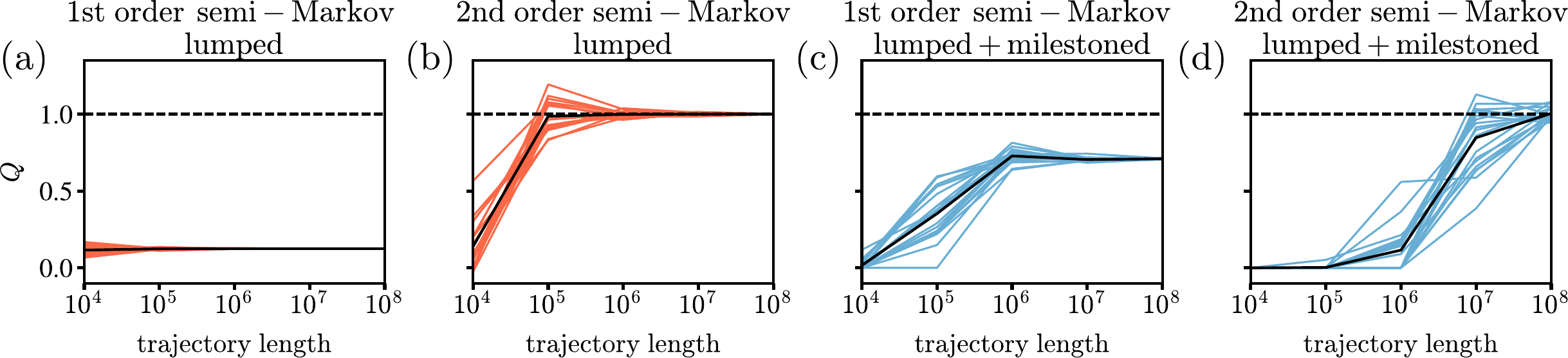}
    \caption{\textbf{Improved thermodynamic inference via milestoning demands better trajectory statistics.} In all panels we consider a uniformly biased single-file system with $N_{\rm vac}=1$, $N_{\rm sites}=9$, $w_{+}=0.46$, and $3$ maximally separated milestones. Colored lines are obtained from $20$ individual trajectories, and the black solid line shows the average. (a-b) Quality factor for the entropy production estimate in the lumped trajectory, based on the $1^{\rm st}$ (a) and $2^{\rm nd}$ (b) order semi-Markov approximation. (c-d) Quality factor for the entropy production estimate in the post-lumped milestoned trajectory, based on the $1^{\rm st}$ (c) and $2^{\rm nd}$ (d) order semi-Markov approximation.}
    \label{S4}
\end{figure}

%---------------------------------
%---------------------------------
In the main manuscript, we pointed out that there is a practical limitation to milestoning a lumped trajectory, which is due to the fact that for increasing milestone distances the estimation of the entropy production requires more sampling. In Fig.~\ref{S4} we show this explicitly for a uniformly biased single-file system with $N_{\rm vac}=1$ and $N_{\rm sites}=9$. Upon comparing the entropy production estimates for the lumped trajectory (Fig.~\ref{S4}a,b) with the milestoned trajectory (Fig.~\ref{S4}c,d), we find that convergence of the latter requires much longer trajectories. This is because larger inter-milestone distances render transitions against the driving increasingly  unlikely. We further observe in Fig.~\ref{S4}b,d that for insufficient sampling, the estimated entropy production can exceed the true entropy production (indicated with the black dashed line) for both lumping and milestoning. Hence, it is always important that sufficiently long (and/or many) trajectories are provided to estimate the entropy production. 

\end{document}